\documentstyle[prd,preprint,aps]{revtex}
\begin{document}
\draft

\noindent
{\bf CAN GENERAL-RELATIVISTIC DESCRIPTION OF GRAVITATION\\ BE
CONSIDERED COMPLETE?}\footnote{Submitted to:  Gravity Research
Foundation (1998); Awarded "Honorable Mention".}
\vskip 0.75in
\noindent
{\bf D. V. Ahluwalia}
\\ Physics Division (P-25), Mail Stop H 846\\
Los Alamos National Laboratory, Los Alamos, New Mexico 87545 USA\\
E-mail: av@p25hp.lanl.gov

\vskip 1.0   in

\begin{quote}
\vskip 0.20in\noindent
The local galactic cluster, the Great attractor, embeds us in a
dimensionless gravitational potential of about $- 3 \times
10^{-5}$. In the solar system this potential is constant to about
$1$ part in $10^{11}$. Consequently, planetary orbits, which are
determined by the gradient in the gravitational potential,  remain
unaffected. However, this is not so for the recently introduced
flavor-oscillation clocks where the  new redshift-inducing phases
depend on the gravitational potential itself. On these grounds, and
by studying the invariance properties of the gravitational
phenomenon in the weak fields, we argue that there exists an
element of incompleteness in the general-relativistic description
of gravitation. An incompleteness-establishing inequality is
derived and an experiment is outlined to test the thesis presented.
\vskip 0.20in
\end{quote}
\newpage
\def\be{\begin{equation}}
\def\ee{\end{equation}}

\noindent
{\bf I.} The gradients of the gravitational potentials are well
known to play a major role in the understanding of motion of the
cosmic bodies. Especially in the weak-field limit of Einstein's
theory of gravitation, they are responsible for the description of,
say, the planetary orbits. In contrast to that, importantly, in the
same limit, there are quantum mechanical effects that depend upon
the gravitational potentials themselves. For example, it was
recently shown that in performing a quantum mechanical linear
superposition of different mass eigenstates  of neutrinos belonging
to different lepton generations, one may create a so-called
``flavor oscillation clock'' that has the remarkable property of
redshifting precisely as required by the Einstein's theory of
gravitation \cite{prd98a}.

In the present study, we demonstrate that such clocks, in principle,
allow to measure the essentially constant gravitational potential
of the local clusters of the galaxies. Taken to its logical
conclusion, this observation results in the question on the
completeness of Einstein's theory of gravitation. In this essay we
systematically explore this question. We come to the conclusion
that, while the gravitationally induced accelerations vanish in a
terrestrial free fall, the gravitationally induced phases of the
flavor-oscillation clocks do not.

We begin with defining the context, then we derive an inequality on
the incompleteness of the general-relativistic description of
gravitation, this is followed by the outline of an experiment to
test the derived inequality, and finally we make some concluding
remarks and summarize the essential thesis of this essay.

\noindent
{\bf II.} As is well known, the solar system is embedded in the
essentially constant gravitational potential of the local cluster
of the galaxies, the so-called Great attractor. This gravitational
potential, denoted by $\Phi_{GA}$ in the following, may be
estimated over the entire solar system to be \cite{K}
\be
\mbox{Solar system:}\quad\Phi_{GA} \equiv \frac{1}{c^2}\,\phi_{GA}
= - 3 \times 10^{-5}.
\ee

For this essay
 the precise value of $\Phi_{GA}$ is not
important, but what is more relevant is that it is {\em constant}
over the entire region of the solar system to an exceedingly large
accuracy of 1 part in $R_{GA-S}/\Delta R_{S}$. Here $\Delta R_{S}$
represents the spatial extent of the solar system, and $R_{GA-S}$
is the distance of the solar system from the Great attractor.
Taking $\Delta R_S$ to be of the order of Pluto's semi-major axis
(i.e., approximately $40$ AU), and $R_{GA-S}$ to be about $40$ Mpc
\cite{K}, we obtain $R_{GA-S}/\Delta R_{S}\sim 10^{11}$. For
comparison, the terrestrial  and solar potentials
on their respective surfaces are of the order
$
\Phi_{E} = - 6.95 \times 10^{-10}$, $\Phi_S = - 2.12 \times 10^{-6}$,
and are therefore much smaller as compared to
$\Phi_{GA}$. Nonetheless, they carry significantly larger gradients over
the relevant experimental regions.

Yet, the constant potential of the Great attractor that pervades the
entire solar system is of no physical consequence within the
general-relativistic context (apart from it being responsible for
the overall local motion of our galaxy). Even the parenthetically
observed motion disappears if we hypothetically and   uniformly
spread the matter of the galactic cluster into a spherical mass to
concentrically surround the Earth. Such a  massive shell in its
interior provides an example  of the gradientless contribution to
the gravitational potential that we have in mind.

A terrestrial freely falling frame that measures accelerations to
an accuracy of less than 1 part in about $10^{11}$ is completely
insensitive to this constant potential. Similarly, since the
planetary orbits are determined by the gradient of the
gravitational potential, they too remain unaffected by this
potential. Nonetheless, in what follows we shall show that quantum
mechanical systems exist that are sensitive to $\Phi_{GA}$. The
simplest example for such a system is constructed in performing
a linear superposition of, say, two different mass eigenstates (see Eqs.
(\ref{Fa}) and (\ref{Fb}) below).

In the next section, $\Phi_{GA}$ shall be considered as a physical
and gradientless gravitational potential as idealized in the
example indicated above. This potential   is to be distinguished
from the usual ``constant of integration'' or the ``potential at
spatial infinity.''

\noindent
{\bf III.} In the following we will exploit the weak-field limit of
gravity as being introduced on experimental grounds. Here, the
phrase ``weak-field limit'' refers to the experimentally
established limit in the weak gravitational fields, rather than to
the limit of a specific theory. Further, although not necessary, for
the sake of the clarity of presentation we shall work in the
non-relativistic domain and neglect any rotation that the
gravitational source may have. This assumption shall be implicit
throughout this essay. The arguments shall be confined to the
system composed of the Earth and the Great attractor, and are
readily extendable to more general situations.

\def\gm{\frac{2G M}{c^2 r}}

For the measurements on Earth the appropriate general-relativistic
(GR) space-time metric is
\be
ds^2 =g^{GR}_{\mu\nu}dx^\mu dx^\nu =\left(1-\gm\right)
dt^2-\left(1+\gm\right) d\bbox r^{\,2},
\label{gmunu}
\ee
where $M$ is the mass of the Earth, $r$ refers to the distance
of the experimental region from Earth's center, and $d\bbox r^{\,2}=
\left(dx^2+dy^2+dz^2\right)$. The
conceptual basis of the theory of general relativity asserts that
the flat space-time metric $\eta^{GR}_{\mu\nu}$,
\be
ds^2 = \eta^{GR}_{\mu\nu} dx^\mu dx^\nu = dt^2 - d\bbox
r^{\,2},\label{etamunu}
\ee
is measured by a freely falling observer on Earth (or, wherever the
observer is). In this framework,  a stationary observer ($\cal O$)
on the Earth may define a gravitational potential according to
\be
\phi_E(\bbox r)= \frac {c^2}{2}\left(g^{GR}_{00}-\eta^{GR}_{00}\right)
=-\frac{c^2}{2}\left(g^{GR}_{\jmath\jmath}-\eta^{GR}_{\jmath\jmath}\right),\quad
\jmath=1,2,3\,\,(\mbox{no sum}).\label{phie}
\ee

One immediately suspects that such a description may not
incorporate the full physical effects of such physical potentials
as $\phi_{GA}$ even though this conclusion is consistent with the
classical wisdom. Indeed,
   the classical equation of motion consistent with the
approximation in Eq. (\ref{gmunu}),
\begin{equation}
m_i\frac{\mbox{d}^2 \bbox r\,} {\mbox{d} t^2} = -\,m_g\bbox\nabla
\phi_E(\bbox r), \label{Newton}
\end{equation}
is invariant under the transformation,
\be
\phi(\bbox r)_E \rightarrow \varphi_E(\bbox r)
=\phi_{GA} + \phi_E(\bbox r). \label{transformation}
\ee
For this reason $\phi_{GA}$ has no apparent effect on the planetary
orbits.

In the quantum realm the appropriate equation of motion is the
Schr\"odinger equation with a gravitational interaction energy term,
\begin{equation}
\left[ -\left(\frac{\hbar^2}{2 m_i}\right)
\bbox\nabla^2 + m_g \phi_{grav}(\bbox r)\right]
\psi(t,\bbox r)
= i\hbar \frac{\partial\psi(t,\bbox r)}{\partial t},\label{Schrodinger}
\end{equation}
as has been confirmed {\em experimentally} in the classic neutron
interferometry experiments of Collela, Overhauser, and Werner
\cite{COW,JJS}. Equation (\ref{Schrodinger})
is not invariant under the
transformation of the type (\ref{transformation}).

Moreover, this lack of invariance does not disappear in
the relativistic   regime where an appropriate relativistic wave
equation, such as the Dirac equation, must be considered. Therefore, the
gravitational potential that appears in Eq. (\ref{Schrodinger})
cannot be identified with $\phi_E(\bbox r)$ of Eq. (\ref{phie}).
To treat the contributions from the Great attractor and the
Earth  on the same footing of physical reality, the
following identification has to be made:
\be
\phi_{grav}(\bbox r) \equiv \varphi_E(\bbox r)
=\phi_{GA} + \phi_E(\bbox r).\label{grav}
\ee

A second observation to be made is to note
that while by setting $m_i=m_g$ in Eq. (\ref{Newton}),
the resulting equation becomes independent of the test-particle
mass; this is {\em not} so for the quantum mechanical equation of
motion (\ref{Schrodinger}) \cite{JJS}.

These two distinctions between the classical and quantum evolutions
lead to the conclusion that the theory of
general relativity for the description of gravitation
cannot be considered complete.
The gravitational potentials as
defined via $g_{\mu\nu}(\bbox r)$ carry an independent physical
significance in the quantum realm, a situation that is reminiscent
on the significance of the gauge potential in electrodynamics
as revealed by the Aharonov-Bohm effect \cite{AB}.

The statement on the general-relativistic incompleteness is best
illustrated on the example of a ``flavor-oscillation clock'' as
introduced in Refs. \cite{prd98a,grf97}. Such clocks are
constructed as a quantum mechanical linear superposition of
different mass eigenstates (for instance, say, two neutrinos from two
different lepton generations \cite{grf96}),
\begin{eqnarray}
\vert F_a\rangle
&&= \cos(\theta) \vert m_1 \rangle + \sin(\theta) \vert m_2
\rangle,\label{Fa} \\
\vert F_b \rangle
&&= -\sin(\theta) \vert m_1 \rangle + \cos(\theta) \vert m_2
\rangle .\label{Fb}
\end{eqnarray}
In the linear superposition of the mass eigenstates we assume ({\em
only} for simplicity) that both $\vert m_1\rangle$ and $\vert
m_2\rangle$ carry vanishingly small three momentum (i.e., are at
rest).

By studying the time oscillation between the flavor states $\vert
F_a\rangle$ and $\vert F_b\rangle$ one discovers that this system
can be characterized by the flavor-oscillation frequency,
\be
\Omega^\infty_{a\rightleftharpoons b}
= \frac{\left(m_2-m_1\right) c^2}{2\hbar}.
\ee

The superscript on $\Omega^\infty_{a\rightleftharpoons b}$ is
to identify  this frequency with a clock at
spatial infinity from the gravitational sources
under consideration (see below).

Now consider this flavor-oscillation clock to be immersed into
the gravitational potential $\varphi_E(r)$.
Then each of the mass eigenstates picks up a {\em
different} phase because the gravitational interaction is of the
form $m \times
\varphi_E(r)$. As a result, one finds that  the new
flavor-oscillation frequency, denoted by $\Omega^{\cal
O}_{a\rightleftharpoons b}$,  is given by \cite{grf97}
\be
\Omega^{\cal O}_{a\rightleftharpoons b} = \left( 1+
\frac{\varphi_E(\bbox r)}{c^2}\right)
\Omega^\infty_{a\rightleftharpoons b}. \label{red}
\ee

This equation is valid for an observer fixed in the global
coordinate system attached to the Earth.

Equation (\ref{red}) would have been the standard gravitational 
redshift expression if the $\varphi_E(\bbox r)$ was replaced by
$\phi_E(\bbox r)$. Freely falling frames $(\cal F)$ do not carry
fastest moving clocks, they carry clocks that are sensitive to
potentials of the type $\phi_{GA}$. A freely falling frame in
Earth's gravity only annuls the gradients of the gravitational
potential while preserving all its constant pieces such as $\phi_{GA}$.
In denoting
by $\Omega^{\cal F}_{a\rightleftharpoons b}$, the frequency as
measured in a freely falling frame on Earth,  one is led to
\be
\Omega^{\cal F}_{a\rightleftharpoons b} = \left( 1+
\frac{\phi_{GA}}{c^2}\right)
\Omega^\infty_{a\rightleftharpoons b}. \label{redb}
\ee
$\,$ From a physical point of view, $\phi_{GA}$ represents contributions
from all cosmic-matter sources. However, all these contributions
carry the same sign. In addition, in the context of the cosmos,
$\Omega^\infty_{a\rightleftharpoons b}$ becomes a purely
theoretical entity. Nevertheless, as shown below,
$\Omega^\infty_{a\rightleftharpoons b}$ does have an operational
meaning.

As a consequence, the following incompleteness-establishing inequality
is found,
\be{\Omega^{\cal F}_{a\rightleftharpoons b}} <
{\Omega^{\infty}_{a\rightleftharpoons b}}.
\ee
This is the primary result of our essay.

\noindent
{\bf IV.} To experimentally test the incompleteness of the
general-relativistic description of gravitation and measure the
essentially constant gravitational potential in the solar system,
we rewrite Eqs. (\ref{red}) and (\ref{redb}) into (to first  order
in the potentials)
\begin{eqnarray}
\frac{\Omega^{\cal O}_{a\rightleftharpoons b}}
{\Omega^{\cal F}_{a\rightleftharpoons b}} && =1+
\frac{\phi_E(\bbox r)}{c^2},
\label{experimenta}\\
\frac{\Omega^{\cal O}_{a\rightleftharpoons b}}
{\Omega^{\infty}_{a\rightleftharpoons b}} &&
=\frac{\phi_{GA}}{c^2}+\left(1+
\frac{\phi_E(\bbox r)}{c^2}\right).\label{experimentb}
\end{eqnarray}
Equation (\ref{experimenta}) shows how the $\phi_{GA}$-dependence
disappears in ${\Omega^{\cal O}_{a\rightleftharpoons b}}/
{\Omega^{\cal F}_{a\rightleftharpoons b}}$. Equation
(\ref{experimentb}), however, indicates that by systematically
measuring ${\Omega^{\cal O}_{a\rightleftharpoons b}}$ as a function
of $\bbox r$, e.g., for an atomic system prepared as a linear
superposition of different energy eigenstates, one can decipher the
existence of $\phi_{GA}$. Because all terrestrial clocks are
influenced by the same $\phi_{GA}$-dependent constant factor, it is
essential that the flavor-oscillation clocks under consideration
integrate the accumulated phase over different paths, thus probing
different $\phi_E(\bbox r)$, and then return to the {\em same}
spatial region in order that all the data interpretation refers to
the same time standard. Such an integration is easily accommodated
in Eq. (\ref{experimentb}). One would then make a  {\em two-parameter fit} 
in $\{{\Omega^{\infty}_{a\rightleftharpoons b}},
\phi_{GA}\}$  to a large set of the
{\em closed-loop integrated} data on $\{{\Omega^{\cal
O}_{a\rightleftharpoons b}(\bbox r)},
\phi_E(\bbox r)\}$. Explicitly,
\be
\oint_\Gamma
{\Omega^{\cal O}_{a\rightleftharpoons b}} (\bbox r) d\ell(\bbox r)
=  {\Omega^{\infty}_{a\rightleftharpoons b}}
\left(1+\frac{\phi_{GA}}{c^2}\right)
\oint_\Gamma d\ell(\bbox r)
+\frac{\Omega^{\infty}_{a\rightleftharpoons b}}{c^2}
\oint_\Gamma{\phi_E(\bbox r)} d\ell(\bbox r)\label{experimentc},
\ee
where $d\ell(\bbox r)$ is the differential length element along the
closed path $\Gamma$. By collecting the data on the ``accumulated
phase'' $\oint_\Gamma {\Omega^{\cal O}_{a\rightleftharpoons b}}
(\bbox r) d\ell(\bbox r)$ and the ``probed gravitational
potential''
$
\oint_\Gamma{\phi_E(\bbox r)} d\ell(\bbox r)
$
for a set of $\Gamma$, and fitting a straight line, one may extract
$\{{\Omega^{\infty}_{a\rightleftharpoons b}},
\phi_{GA}\}$. Rigorously speaking, what
one obtains  is  ${\Omega^{\infty}_{a\rightleftharpoons b}}$ and
the constant $\phi_{GA}$ as modified by other cosmic contributions.
Further, these additional contributions may include extra
general-relativistic contributions from the yet-unknown
interactions that may couple to the various parameters associated
with the superimposed quantum states.

A simple consideration of the magnitude of various gravitational
potentials involved and the accuracy of clocks based on quantum
superpositions of atomic states leads to the tentative conclusion
that the suggested experiment is feasible within the existing
technology. In this regard note is taken that various ionic and
atomic clocks have reached an accuracy of $1$ part in $10^{15}$
with a remarkable long-term stability. In addition, workers in this
field are optimistic that a several-orders-of-magnitude improvement
may be expected in the next few years (see, e.g, Barbara  Levi's
recent coverage of this subject in the February 1998 issue of
{\em Physics Today} \cite{clocks}).

\noindent
{\bf V.} In the present study we emphasized observability of the
constant potential of the Great attractor by means of flavor-oscillation 
clocks. While in a classical context, the force $\bbox
F = - m_g \bbox
\nabla \phi (\bbox r)$ experienced by an object is independent of
gradientless gravitational potentials such as $\phi_{GA}$;
the frequency of the flavor oscillation clocks
depends directly on $\phi_{GA}$ [in addition to
$\phi_E(\bbox r)$].

The above considerations suggest that in a free fall the space-time
interval is given by
\def\gmf{\frac{2\phi_{GA}}{c^2 }}
\be
{\cal F}:\quad ds^2_{\cal F}  =\eta^{\cal F}_{\mu\nu} dx^\mu
dx^\nu=
\left(1+\gmf\right)
dt^2-\left(1-\gmf\right)d\bbox r^{\,2},
\label{chimunu}
\ee
and not by Eq. (\ref{etamunu}), as asserted by the foundations of
the theory of general relativity. Simultaneously, Eq. (\ref{gmunu})
is to be replaced by
\def\gmn{\frac{2\varphi_E(\bbox r)}{c^2}}
\be
{\cal O}:\quad
 ds^2_{\cal O} =g_{\mu\nu}^{\cal O} dx^\mu
dx^\nu=\left(1+\gmn\right) dt^2-\left(1-\gmn\right) d\bbox r^{\,2},
\ee
with Eq. (\ref{etamunu}) now remaining valid  only at the ``spatial
infinity''
\be
\infty:\quad
ds^2_{\infty} = \eta^\infty_{\mu\nu} dx^\mu dx^\nu = dt^2 - d\bbox
r^{\,2}.\label{etamunuinfty}
\ee
The symbols $\{{\cal F}:,\,\,{\cal O}:,\,\,\infty :\}$ in the above
equations are to remind the reader of  the related observers. These
equations are expected to hold at least in the quantum realm.

Such  modifications are perfectly justified because of the
linearity of the weak-field limit, where one is able to formulate
the physics in terms of the additive gravitational potentials.

Within the considered framework and approximations, the space-time
curvatures derived from $g^{GR}_{\mu\nu}$ and $g^{\cal O}_{\mu\nu}$
are identical. A similar statement applies to $\eta^{GR}_{\mu\nu}$
and $\eta^{\cal F}_{\mu\nu}$. Yet, the quantum effects of
gravitation do not vanish in a freely falling frame, they vanish at
spatial infinity. Consequently, the observable  gravitational
potential (as detected, e.g.,  by the flavor-oscillation clocks) is
given by
\be\phi_{grav}(\bbox r)=
\frac{c^2}{2}\left(g^{\cal O}_{00}(\bbox r)-\eta^\infty_{00}\right).
\ee
This result is in agreement with  Eq. (\ref{grav}) and differs from
the general-relativistic result contained in Eq. (\ref{phie}). The
theory of general relativity implicitly assumes equality of
$\eta_{\mu\nu}^\infty$ and $\eta_{\mu\nu}^{\cal F}$, and thereby
omits physical effects of the gradientless physical potentials in
its treatment of the freely falling observers. Here, by examining
the classical and quantum realm in the weak gravitational fields
(where the difficulties of ``quantum gravity'' are avoided), we
have shown that this implicit general-relativistic assumption has
to be abandoned, because while
$\eta_{\mu\nu}^{\infty}=\eta_{\mu\nu}^{GR}$, $\eta_{\mu\nu}^{\cal
F}\ne\eta_{\mu\nu}^{GR}$. That is, $\eta_{\mu\nu}^{\cal F}$ is to
be physically distinguished from $\eta_{\mu\nu}^{\infty}$.

The reported incompleteness in the theory of general relativity for
the description of gravitation also reveals certain similarities to
the Aharonov-Bohm effect \cite{AB}. Indeed, in the Aharonov-Bohm
effect an observable phase arises in a region with vanishing field
strength tensor $F^{\mu\nu}(\bbox r)$, (i.e., in a region with
vanishing $4$-curl of the gauge potential $A^{\mu}(\bbox r)$). In
the effect reported here,  an observable phase arises in a region
where the contributions of the $\phi_{GA}$-type constant potentials
to the curvature tensor $R^{\mu\nu\sigma\lambda}(\bbox r)$ vanish.
Both of the effects mentioned above illustrate the circumstance
that in quantum mechanical processes the gauge potential
$A^\mu(\bbox r)$ and the gravitational potential $g^{\mu\nu}(\bbox
r)$ may be favored over the corresponding fields strength tensor
$F^{\mu\nu}(\bbox r)$, and the curvature tensor
$R^{\mu\nu\sigma\lambda}(\bbox r)$, respectively.

However, since the number of the independent degrees of freedom
of $A^\mu(\bbox r)$ is quite different from that of $g^{\mu\nu}(\bbox r)$,
the  analogy between the Aharonov-Bohm effect
and the one considered here is not complete.

In summary, the local galactic cluster, the Great attractor, embeds
us in a dimensionless gravitational potential of about $- 3 \times
10^{-5}$. In the solar system this potential is constant to about
$1$ part in $10^{11}$. Consequently, planetary orbits remain
unaffected. However, this is not so for the flavor-oscillation
clocks. In a terrestrial free fall  the gravitationally induced
accelerations vanish, but the gravitationally induced phases
of the flavor-oscillation clocks do not. We argue that
there exists an element of incompleteness in the
general-relativistic description of gravitation. The arrived
incompleteness may be subjected to an experimental test by
verifying the inequality derived here.

The origin of the reported incompleteness  lies in the implicit
general-relativistic assumption on  the equivalence of the
space-time metric as measured by a freely falling observer in the
vicinity of a gravitating source  (which in turn is embedded in a
$\Phi_{GA}$-type constant gravitational potential)
 and the space-time metric as
measured by an observer at ``spatial infinity.''

\medskip
\noindent{\bf ACKNOWLEDGMENTS}
\bigskip

I feel obliged to record a suggestion of Sam Werner. In a
conversation in Missouri, about a year ago, he brought to my
attention that he wishes to surround one of the arms of a neutron
interferometer in a hollow cylinder filled with about a ton (if I
remember correctly) of mercury. Thus, effects of a constant
gravitational potential could be experimentally studied in a
neutron interferometer. It appears that by appropriately varying
the amount of mercury in the hollow cylinder one may also address
the incompleteness issue of the theory of general relativity. It is
also my pleasure to thank V. Raatriswapan for many useful comments.
This work was done, in part, under the auspices of the U.S.
Department of Energy.

\end{document}